\documentclass[12pt,a4paper]{article}
\usepackage{setspace}
\usepackage{epsfig}
\begin{document}

\title{\bf\large The $\phi\to\eta\pi^0\gamma$ decay }

\author{\normalsize M.N.Achasov, S.E.Baru, K.I.Beloborodov, 
A.V.Berdyugin, A.V.Bozhenok, \\
\normalsize 
A.D.Bukin, D.A.Bukin, S.V.Burdin, T.V.Dimova, S.I.Dolinsky, V.P.Druzhinin, \\
\normalsize M.S.Dubrovin, I.A.Gaponenko, V.B.Golubev, V.N.Ivanchenko,
A.A.Korol, \\
\normalsize  M.S.Korostelev, 
S.V.Koshuba, A.A.Mamutkin,
\underline{E.V.Pakhtusova\thanks{e-mail:pakhtusova@inp.nsk.su; FAX:
+7(3832)342163}},
E.E.Pyata, \\ 
\normalsize A.A.Salnikov, S.I.Serednyakov, V.V.Shary, Yu.M.Shatunov,
V.A.Sidorov, \\
\normalsize Z.K.Silagadze, A.N.Skrinsky,  A.V.Vasiljev \\ \\
\normalsize Budker Institute of Nuclear Physics, Novosibirsk State 
University,\\
\normalsize Novosibirsk, 630090, Russia}
\date{}
\maketitle

\begin{abstract}
\normalsize
Rare radiative decay $\phi\to\eta\pi^0\gamma$ was studied
with SND detector at  \mbox{VEPP-2M} electron-positron
collider and its branching ratio was measured:
$B(\phi\to\eta\pi^0\gamma)=(0.88\pm0.14\pm0.09)\cdot 10^{-4}$.
Significant  contribution of the $a_0(980)\gamma$ intermediate 
state was observed in the decay. The result is based on total
integrated luminosity corresponding to $2\cdot10^7$ produced
$\phi$ mesons. 
\\
\\
{\it PACS:} 13.25.-k; 13.65.+i; 14.40.-n\\
{\it Keywords:} $e^+e^-$ collisions; Vector meson; Detector
\end{abstract}

\twocolumn

{\large\bf Introduction.}
\normalsize 
The first observation of the rare radiative  decay   
\begin{equation}
\label{a0g1}
          \phi \to \eta\pi^0\gamma,
\end{equation}
and measurement of its branching ratio were performed in
Novosibirsk by SND detector at VEPP-2M $e^+e^-$ collider 
\cite{ca0g1, ca0g2}. The analysis was based on
data collected by SND in 1996 \cite{SND1}. Later the results were
confirmed by CMD-2 group in their recent publication \cite{CMD1}.
Results of the present work are based on analysis of all SND data collected in the
vicinity of $\phi$ meson in 1996--1998 \cite{SND2}. 

Reaction (\ref{a0g1}) is especially interesting in connection with
the scalar $a_0(980)$ meson problem, which is being actively discussed
in the literature. At present there is no
generally accepted  viewpoint on the nature of $a_0$,
its quark structure is still not well established, and several models exist
including modification of $q\bar{q}$ scheme \cite{theor1},
$ K\bar{K}$ molecular model \cite{theor2}, and
4-quark model \cite{theor3}.
It was suggested in \cite{theor4} that the decay
$\phi \to a_0(980)\gamma \to \eta\pi^0\gamma $ may
serve as a probe of the $a_0$-meson quark structure.
Theoretical predictions for the decay branching ratio  vary from 
$10^{-5}$ for the simple two-quark and $K\bar{K}$
molecular models up to $10^{-4}$ in
the 4-quark model \cite{theor4, theor5}. 
There also exist some models in which high
rate of the decay (\ref{a0g1}) can be achieved without assumption of
4-quark structure of the $a_0$ meson \cite{theor6}.
Another possible mechanism 
of the $\phi \to \eta\pi^0\gamma $ decay
is $\phi\to \rho^0\pi^0, \rho\to\eta\gamma $. 
The vector meson dominance model prediction for this branching
ratio is $5\cdot10^{-6}$  \cite{theor4,theor5}.
Detailed study of the $\phi \to \eta\pi^0\gamma $ 
decay may provide decisive information on $a_0(980)$ meson problem.

{\large\bf Experiment.}
\normalsize 
The  SND \cite{SND}
is a universal nonmagnetic detector. Its main part  
is  a 3-layer electromagnetic calorimeter consisted of 1630 NaI(Tl)
crystals. The energy resolution of the calorimeter for photons can be
described as
$\sigma_E/E=4.2\%/\sqrt[4]{E(GeV)}$ \cite{calor}, the angular resolution 
is close to $1.5^\circ$. The solid angle coverage is 
$90\%$ of $4\pi$ steradian.

The data used for the study of $\phi \to \eta\pi^0\gamma$ decay
were collected in 1996-1998 \cite{SND2}.
Nine successive scans of the energy range 980--1040~MeV were performed.
The data were collected at 16 beam energy points.
The total integrated luminosity in the experiment is 12 $pb^{-1}$ 
and total number of produced $\phi$ mesons --- $2\cdot10^7$.

{\large\bf Event Selection.} 
 Main sources of background for the  process under study
\begin{equation}
 e^+e^- \to \phi \to \eta \pi^0 \gamma \to 5\gamma \label{etapi0g}
\end{equation}
are the following $\phi$-meson decays:   
\begin{equation}
 e^+e^- \to \phi \to \pi^0 \pi^0 \gamma \to 5\gamma \label{pi0pi0g}
\end{equation}
\begin{equation}
 e^+e^- \to \phi \to \eta \gamma \to 3\pi^0\gamma \to 7\gamma \label{etag},
\end{equation}
\begin{equation}
 e^+e^- \to \phi \to K_S K_L \to neutrals \label{kskl}
\end{equation}
and a nonresonant process
\begin{equation}
 e^+e^- \to \omega\pi^0 \to \pi^0 \pi^0 \gamma \to 5\gamma \label{ompi}.
\end{equation}

The process (\ref{etag}) does not produce 5$\gamma$ events directly but
can mimic the process (\ref{etapi0g})
due to either merging of
close photons or loss of soft photons through openings in the calorimeter.
The process (\ref{kskl}) contributes due to 
$K_S\to\pi^0\pi^0$ decay accompanied by either nuclear interaction of
the $K_L$ meson or its decay in flight.

Primary event selection was based on simple criteria: the number of
reconstructed photons is equal to five, there are no tracks in the
drift chamber, the total energy deposition $E_{tot}$
ranges from 0.8 up to 1.1 of the center of mass energy $2E_0$,
the total transverse momentum of photons
is less than $0.15E_{tot}/c$. In order to suppress background from the
processes (\ref{etag}) and (\ref{kskl}) a special ``photon quality''
parameter $\zeta$ \cite{phqu} was used.
For \mbox{$i$-th} reconstructed photon
the ${\zeta}_i$ is a minus logarithm of likelihood for
the corresponding transverse energy deposition profile
observed in the calorimeter to be produced by a single photon.
For multiphoton events $\zeta$ is defined as a maximum ${\zeta}_i$.
The cut $\zeta < 0$ suppresses the background 
from the process (\ref{etag}) by a factor of two, reducing the detection
efficiency for actual 5-$\gamma$ events by only 8\%.
To suppress beam background photons which appear
mostly in the calorimeter areas closest to the beam and are relatively soft,
additional cut was imposed on polar angles of the two softest photons in an event:
$32^\circ<\theta_{4},\theta_{5}<148^\circ $.

For events which  passed the cuts described above 
kinematic fitting under two alternative hypotheses was performed
and corresponding values of $\chi^2$ calculated:  
\begin{itemize}
\item an event is one of the process  $e^+e^- \to 5\gamma$;
the $\chi^2$ value is denoted as $\chi_{5\gamma}^2$;
\item an event is one of the process  $e^+e^- \to 3\gamma$ with two
additional stray photons;
      the $\chi^2$ value is denoted as $\chi_{3\gamma}^2$.
\end{itemize}

The following restrictions on the
$\chi _ {5\gamma}^2$ and $\chi _ {3\gamma}^2$ parameters were imposed: 
$\chi_{5\gamma}^2 < 25, \, \chi_{3\gamma}^2> 20.$
The first restriction causing only 5\%
loss of actual $\phi \to \eta\pi^0\gamma $ events reduces background from 
the process (\ref{etag}) by
approximately $30\%$  and almost completely removes 
background from the  
process (\ref{kskl}). The second cut suppresses
background from the processes $\phi\to\eta\gamma\to 3\gamma$,
$\phi\to\pi^0 \gamma\to 3\gamma$, $e^+e^- \to 2\gamma,3\gamma$ (QED). 

For further background suppression, an event configuration
(the photon energies and  angles after 5-$\gamma$ kinematic fitting)
was compared using modification of kinematic fitting technique developed
in the work
\cite{burot} with ones expected for the process
(\ref{etapi0g}) and background processes (\ref{pi0pi0g}),
(\ref{ompi}). Corresponding measure of discrepancy $P$ is an increase
of $\chi^2$ for a 5-$\gamma$ event after application of additional
requirements on intermediate states for each tested hypothesis.
The following hypotheses were considered:        
\begin{itemize}
\item
an event is a cascade reaction 
$e^+e^- \to X\gamma,\; X\to\eta\pi^0$ where $X$ is some
intermediate particle; the $P_{\eta\pi\gamma}$ parameter
and  invariant masses of photon pairs, presumably produced in
the decays of $\pi^0$  and $\eta$  mesons ( $M_{\pi}$ and $M_{\eta}$ )
were calculated;
\item
an event is a cascade reaction
$e^+e^- \to X\gamma,\; X\to\pi^0\pi^0$;
the $P_{\pi\pi\gamma}$ parameter was calculated;
\item 
an event is of the process $e^+e^-\to\omega\pi^0, \omega\to
\pi^0\gamma$;  parameter $M_{\omega}$ --- an
invariant mass of $\pi^0\gamma$  pair from the $\omega \to \pi^0\gamma$ decay, 
was calculated;
\end{itemize}

Relative contributions from the background processes (\ref{pi0pi0g}) and
(\ref{ompi}) vary
with $m_{\eta\pi}$ --- invariant mass of $\eta\pi^0$ pair.
At $m_{\eta\pi} < 975$~MeV the
background from the process (\ref{pi0pi0g}) becomes significant.
Additional cut $P_{\pi\pi\gamma} > 2$ suppresses it by a factor of
three reducing detection efficiency for the process (\ref{etapi0g})
by only $12\%$.
At $m_{\eta\pi} \leq 900$~MeV the dominant background comes
from the process (\ref{ompi}). In this region
restriction $M_{\omega} < 725$~MeV removes background almost completely.

\begin{figure}[htb]
   \epsfig{figure=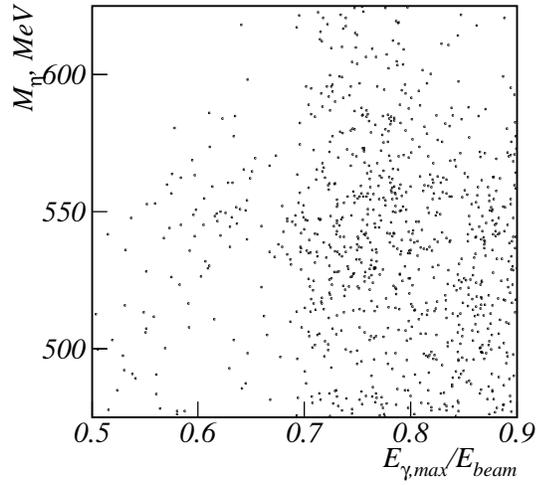,width=0.45\textwidth}
   \caption{ \label{amefer1n}
Distribution of $M_{\eta}$ --- reconstructed mass of $\eta$
meson versus the energy of the most energetic photon 
in the event $E_{\gamma \, max}/E_{beam}$. 
}
\end{figure}

\begin{figure}[htb]
   \epsfig{figure=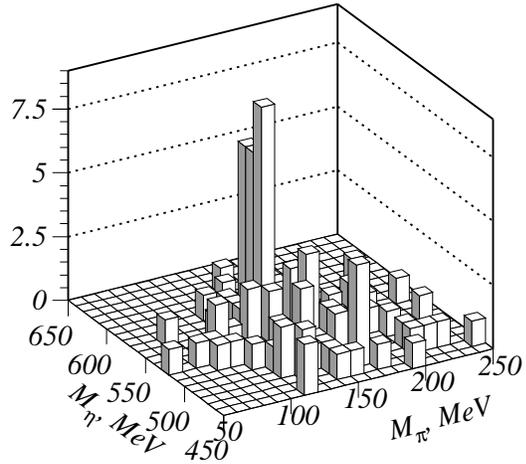,width=0.45\textwidth}
   \caption{ \label{meta}
Distribution of $M_{\eta}$ --- reconstructed mass of $\eta$
meson versus  $M_{\pi}$ --- reconstructed mass of $\pi^0$ 
for events with
$ E_{\gamma \, max}/E_{beam} < 0.68$. 
}
\end{figure}

The scatter plot for $M_{\eta}$ invariant mass
versus $E_{\gamma \, max}/E_{beam}$ --- normalized energy of the most 
energetic photon
in the selected events is shown in fig.\ref{amefer1n}, where
two regions are distinguishable: \\
$E_{\gamma \, max}/E_{beam} > 0.68$ dominated by
background from the reaction (\ref{etag}) with a nearly uniform  $M_{\eta}$
distribution and \\ 
$E_{\gamma \, max}/E_{beam}<0.68$
where the background is small and the points are grouped close to
$\eta$-meson mass.
The $M_{\eta}$ vs. $M_{\pi^0}$ distribution in events with
$E_{\gamma \, max}/E_{beam} < 0.68$ is shown in
fig.\ref{meta}. It is clearly peaked at $\pi^0$ and $\eta$-meson masses.

\begin{figure}[htb]
   \epsfig{figure=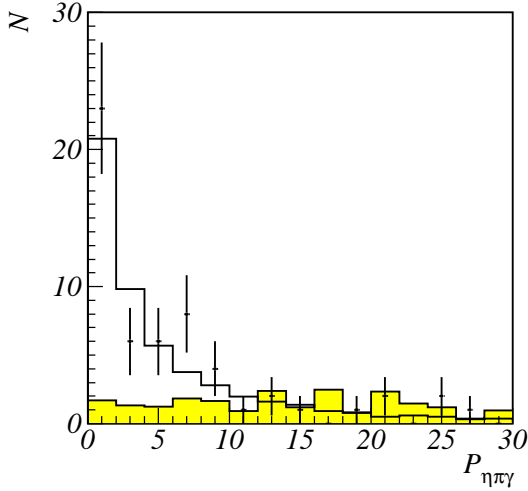,width=0.45\textwidth}
   \caption{ \label{pepg}
 The $ P_{\eta\pi\gamma}$ distribution.
 Points with error bars - experimental data.
 Histogram --- simulated signal from $\phi\to\eta\pi^0\gamma$ decay
 corresponding to branching ratio of
 $0.9\cdot10^{-4}$,
shaded histogram --- estimated  background from the 
$e^+e^- \to \omega\pi^0$
and $\phi \to \eta\gamma, f_0(980)\gamma$ processes.
}
\end{figure}

For final selection   of $\phi \to \eta\pi\gamma$ events,
in addition to the cuts described above
the restriction  $ E_{\gamma \, max}/E_{beam} < 0.68$
practically completely removing background from the
process (\ref{etag}) and $ P_{\eta\pi\gamma} < 7$ were applied.
The $P_{\eta\pi\gamma}$ distribution for the selected
events is shown in fig.\ref{pepg}.
The cut $P_{\eta\pi\gamma} < 7$ roughly corresponds to the
restrictions
$ |M_{\pi^0} -m_{\pi^0} | < 30$~MeV and
$ |M_{\eta} -m_{\eta} | < 30$~MeV,
where $m_{\pi^0}$ and $m_{\eta}$ are the $\pi^0$ and $\eta$ masses.
Total of 39 events were found with expected background of
$3.2\pm0.7$ events. In the region $7 < P_{\eta\pi\gamma} <15$
ten events were found in agreement with estimated $8.9\pm0.4$
$\eta\pi^0\gamma$ events plus $ 6\pm1$  background events.
Thus, after cuts described above the event sample
still contains background of about $10\%$.
After background subtraction  $35.8\pm6.3$ events of the process
$e^+e^- \to \eta\pi^0\gamma$ are left.

\begin{figure}[htb]
   \epsfig{figure=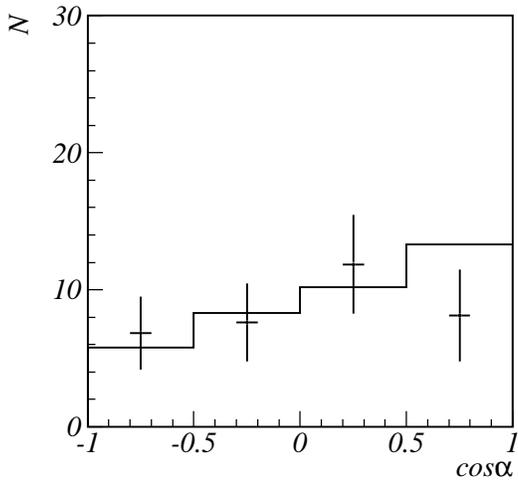,width=0.45\textwidth}
   \caption{ \label{cosa}
The $\cos\alpha$ distribution. $\alpha$ is an angle between
recoil photon and $\eta$ meson in the $\eta\pi^0$ rest frame for
selected $\eta\pi^0\gamma$ events.
Points with error bars - experimental data, histogram - simulation of
the process (\ref{a0g1}) with $BR=0.9\cdot10^{-4}$.
}
\end{figure}

\begin{figure}[htb]
   \epsfig{figure=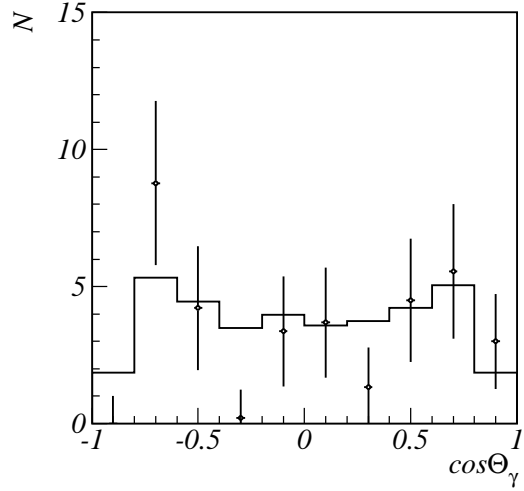,width=0.45\textwidth}
   \caption{ \label{cosg}
Recoil photon polar angle distribution for selected $\eta\pi^0\gamma$ events.
Points with error bars - experimental data, histogram - simulation of
the process (\ref{a0g1}) with $BR=0.9\cdot10^{-4}$.
}
\end{figure}

{\large\bf Data analysis.}
Fig.\ref{cosa} shows $\cos\alpha$ distribution for
the selected events, where $\alpha$ is
an angle  between the recoil photon in the reaction
(\ref{etapi0g})
and $\eta$-meson momentum in the $\eta\pi^0$ rest frame.
Estimated background of 3.2 events is subtracted.
Experimental distribution is in a good agreement with the simulated
one, which was initially isotropic as expected for a scalar
intermediate state. Its visible slope is a consequence of the
$E_{\gamma \, max}< 0.68$ cut.
Such an agreement may be considered as an evidence that
$\eta\pi^0$-system is produced in a scalar state ($P(\chi^2) = 61\%$). 
In fig.\ref{cosg} the $\cos\theta_{\gamma}$ distribution is shown.
The $\theta_{\gamma}$ is a polar angle of the recoil photon in the
reaction (\ref{etapi0g}). It  also agrees ($P(\chi^2) = 22\%$)
with the simulated distribution
$(1+\cos^2\theta_{\gamma})$ expected for production of
a scalar particle and photon.

Detection efficiency obtained by simulation must be corrected
for event loss due to additional spurious photons and for 
imprecise simulation of parameters used in the event
selection cuts. Corresponding  correction
factor was obtained from experimental data. To this end
cross section of the process (\ref{ompi}) was measured using the 
selection criteria similar to those described above for the process 
(\ref{etapi0g}). The result was compared with our earlier measurement 
\cite{ompid}. It was found that simulation overestimates
detection efficiency for the 
process (\ref{etapi0g}) by $5\%$.

In the Table~\ref{tab}  the numbers of selected events,
detection efficiencies, and measured differential branching ratios
$ dB/dm $  as a function of $m_{\eta\pi}$ invariant mass are listed.
The detection
efficiencies and $dB/dm$ values are given at middle points of the corresponding
invariant mass bins. Uniformly distributed background of 3 events 
was subtracted.
The detection efficiency 
averaged over the experimental invariant mass spectrum is equal to $2.1\%$.

Fitting of energy dependence of the experimental cross section
 by the sum of the resonant cross section of the process
($\ref{etapi0g}$) and energy-independent background
results in 
\begin{equation}
\label{res2}
B(\phi\to\eta\pi^0\gamma)=(0.88\pm0.14\pm0.09)\cdot10^{-4},
\end{equation}
The main sources of systematic error here are uncertainties in
the measured cross section of the process (\ref{ompi}) and in average
detection efficiency due to large statisical error of the 
observed $\eta\pi^0$ invariant mass spectrum. 

\begin{figure}[htb]
   \epsfig{figure=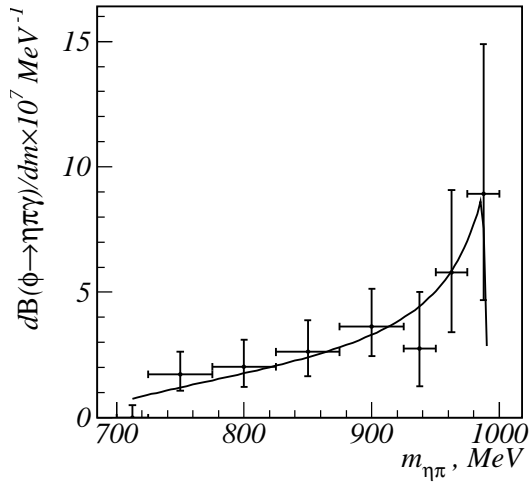,width=0.45\textwidth}
   \caption{ \label{metap}
$ M_{\eta\pi}$ invariant mass spectrum.
The fitted curve corresponds to
$M_{a_0} = 995^{+52}_{-10}$~MeV 
$g^{2}_{a_0K^{+}K^{-}}/4\pi   = (1.4^{+9.4}_{-0.9})\mbox{GeV}^2$ 
$g^2_{a_0\eta\pi}/4\pi = (0.77^{+1.29}_{-0.20})\mbox{GeV}^2$
}
\end{figure}

In fig.\ref{metap} the dependence of the measured
$\phi \to \eta\pi^0\gamma$ decay branching ratio
on the invariant mass of the $\eta\pi^0$ pair is shown.
In spite of smaller recoil photon phase space at
high $\eta\pi^0$ invariant masses the observed mass
spectrum shows enhancement in this region. This means
that $\eta\pi^0$ system is produced in some resonant
state. The only known resonance which have relevant
mass and quantum numbers is $a_0(980)$ and
the observed enhancement at large $\eta\pi^0$
invariant masses  can be described as
manifestation of $\phi\to a_0\gamma$ decay.
Known $\phi\to \rho^0\pi^0, \rho\to\eta\gamma $ 
decay mechanism must produce $\eta\pi^0$ pairs with smaller
invariant masses and as was already mentioned its branching
ratio is much smaller than observed one,
although its amplitude should be taken into
account in the approximation of the whole mass spectrum
in future high statistics experiments.
For $M_{\eta\pi} > 900$~MeV we have: 
\begin{equation}
\label{res3}
B(\phi\to\eta\pi^0\gamma)=(0.46\pm0.13)\cdot10^{-4}.
\end{equation}

{\large\bf Discussion.}
Since the experimental data show large contribution from the 
$ \phi\to a_0\gamma $ decay an attempt was made to approximate
observed invariant mass spectrum in assumption of pure 
$\phi \to a_0\gamma$ and assuming decay
dynamics as described in the work \cite{theor4}. 
This hypothesis gives rather good approximation of the experimental data.
The fitting curve is shown in 
fig.\ref{metap} ($P(\chi^2) = 61\%$). The following optimal values of the
$a_0$-meson parameters  were obtained:
\begin{equation}
\begin{array}{l}
   M_{a_0} = 995^{+52}_{-10} \mbox{MeV} \\
   g^{2}_{a_0K^{+}K^{-}}/4\pi   = (1.4^{+9.4}_{-0.9})\mbox{GeV}^{2} \\
   g^2_{a_0\eta\pi}/4\pi = (0.77^{+1.29}_{-0.20})\mbox{GeV}^{2}
\end{array}
\end{equation}
The corresponding fitting curve is shown in fig.\ref{metap}.
The optimum value of the ratio 
$g_{a_0\eta\pi}/g_{a_0K^{+}K^{-}}=0.75^{+0.52}_{-0.32}$ 
within experimental
errors
satisfies the relation between coupling constants 
$g_{a_0\eta\pi}=\sqrt{2/3}\cdot g_{a_0K^{+}K^{-}}$ obtained in \cite{theor3}
in the assumption of 4-quark structure of $a_0$ meson.
If we fix this ratio at its 4-quark model prediction,
the optimal values of other fit parameters become:          
\begin{equation}
\begin{array}{l}
   M_{a_0} = 994^{+33}_{-8} \mbox{MeV} \\
   g^{2}_{a_0K^{+}K^{-}}/4\pi   = (1.05^{+0.36}_{-0.25})\mbox{GeV}^{2} \\
\end{array}
\end{equation}
The mass  $ M_{a_0}$ is in agreement with the PDG value $983.4$~MeV \cite{PDG}.
 
Comparison with (\ref{res3}) shows that about 50\%
of the observed branching
ratio (\ref{res2}) corresponds to $M_{\eta\pi} > 900$~MeV and
the observed invariant mass spectrum is consistent with
the model \cite{theor4}. Thus it can be assumed, that the
$a_0\gamma$ intermediate state dominates in this decay. Other
decay mechanisms may contribute, for example 
$\phi \to \rho^0\pi^0, \rho\to\eta\gamma $, although rough estimation
shows, that its contribution can be neglected at present level of experimental
errors (\ref{res2}). Assuming pure $\phi\to a_0 \gamma$ decay, we get
\begin{equation}
B(\phi \to a_0\gamma)=(0.88\pm0.17)\cdot10^{-4}.
\end{equation}

{\large\bf Conclusions.}
 In this work, using experimental data corresponding to about
$2\cdot10^7$ produced $\phi$ mesons,   36 events of  
the $ \phi\to \eta\pi^0\gamma $ decay were 
found. The measured   branching ratio  of this decay
$ B(\phi\to\eta\pi^0\gamma) =(0.88\pm0.14\pm0.09)\cdot10^{-4}$
is in agreement with our previous result 
$ B(\phi\to\eta\pi^0\gamma) =(0.83\pm0.23)\cdot10^{-4}$
\cite{ca0g2}, based on analysis of a 
part of the experimental statistics, as well as
with the CMD-2 measurement: 
$ B(\phi\to\eta\pi^0\gamma) =(0.90\pm0.24\pm0.10)\cdot10^{-4}$
\cite{CMD1}. 
Observed enhancement in the $\eta\pi^0$-pair invariant mass spectrum  
at large masses shows large contribution of
$a_0\gamma$ intermediate state.
Assuming dominance of this mechanism we obtain
$ B(\phi \to a_0\gamma) = (0.88\pm0.17)\cdot10^{-4}$.             

{\large\bf Acknowledgement.}
    The authors express their gratitude to N.N.Achasov
    for fruitful discussions.
    
This work is supported in part by ``Russian Fund for Basic 
Researches'' (Grant No. 99-02-16813), 
``Russia Universities'' Fund (Grant No. 3H-339-00) and 
STP ``Integration'' Fund (Grant No 274).

\pagebreak
\begin{center}
{\bf Figure captions}
\end{center}
\begin{list}{\textbullet}{}
\item{\it Figure} \ref{amefer1n}: 
Distribution of $M_{\eta}$ --- reconstructed mass of $\eta$
meson versus the energy of the most energetic photon 
in the event $E_{\gamma \, max}/E_{beam}$. 

\item{\it Figure} \ref{meta}:
Distribution of $M_{\eta}$ --- reconstructed mass of $\eta$
meson versus  $M_{\pi}$ --- reconstructed mass of $\pi^0$ 
for events with
$ E_{\gamma \, max}/E_{beam} < 0.68$. 

\item{\it Figure} \ref{pepg}
 The $ P_{\eta\pi\gamma}$ distribution.
 Points with error bars - experimental data.
 Histogram --- simulated signal from $\phi\to\eta\pi^0\gamma$ decay
 corresponding to branching ratio of
 $0.9\cdot10^{-4}$,
shaded histogram --- estimated  background from the 
$e^+e^- \to \omega\pi^0$
and $\phi \to \eta\gamma, f_0(980)\gamma$ processes.

\item{\it Figure} \ref{cosa}
The $\cos\alpha$ distribution. $\alpha$ is an angle between
recoil photon and $\eta$ meson in the $\eta\pi^0$ rest frame for
selected $\eta\pi^0\gamma$ events.
Points with error bars - experimental data, histogram - simulation of
the process (\ref{a0g1}) with $BR=0.9\cdot10^{-4}$.

\item{\it Figure} \ref{cosg}
Recoil photon polar angle distribution for selected $\eta\pi^0\gamma$ events.
Points with error bars - experimental data, histogram - simulation of
the process (\ref{a0g1}) with $BR=0.9\cdot10^{-4}$.

\item{\it Figure} \ref{metap}
$ M_{\eta\pi}$ invariant mass spectrum.
The fitted curve corresponds to
$   M_{a_0} = 995^{+52}_{-10}$~MeV 
$   g^{2}_{a_0K^{+}K^{-}}/4\pi   = (1.4^{+9.4}_{-0.9})\mbox{GeV}^2$
$   g^2_{a_0\eta\pi}/4\pi = (0.77^{+1.29}_{-0.20})\mbox{GeV}^2$
\end{list}

\pagebreak

\clearpage
\pagebreak
\newpage

\begin{table}
\begin{center}
\caption{\label{tab}
 The number  $N$ of found $\phi\to\eta\pi^0\gamma$ decay
 events,
 detection efficiency $\epsilon$, and measured $dB/dm$ 
 as a function of $m_{\eta\pi}$ --- the $\eta\pi^0$ invariant mass.   
}
\begin{tabular}{|c|c|c|c|}
\hline
  & & &  \\
 $m_{\eta\pi},MeV$ & $N$ &$\epsilon$ &$dB/dm\cdot10^7 $ \\
  & & & ($MeV^{-1}$)\\
\hline
700-725 & 0 & 0.039 & $0 + 0.5$ \\
725-775 & 5 & 0.029 & $1.7^{+0.9}_{-0.7}$ \\
775-825 & 6 & 0.026 & $2.0^{+1.1}_{-0.8}$ \\
825-875 & 7 & 0.024 & $2.6^{+1.3}_{-1.0}$ \\
875-925 & 9 & 0.023 & $3.6^{+1.5}_{-1.2}$ \\
925-950 & 3 & 0.018 & $2.8^{+2.3}_{-1.5}$ \\
950-975 & 5 & 0.016 & $5.8^{+3.3}_{-2.4}$ \\
975-1000 & 4& 0.011 & $8.9^{+6.0}_{-4.2}$ \\
\hline
\end{tabular}
\end{center}
\end{table}

\end{document}